\documentclass[prl,nofootinbib,twocolumn]{revtex4}
\def\mysection#1{{\bf #1.} }
\def\mysections#1{{\bf #1.} }

\usepackage{amssymb}
\usepackage{amsmath}
\usepackage[dvips]{graphicx}
\usepackage{longtable}
\usepackage{verbatim}
\usepackage{amsfonts}

\newcommand{\be}{\begin{equation}}
\newcommand{\ee}{\end{equation}}
\newcommand{\bea}{\begin{eqnarray}}
\newcommand{\eea}{\end{eqnarray}}
\newcommand{\beq}{\begin{equation}}
\newcommand{\eeq}{\end{equation}}
\def\beqa{\begin{eqnarray}}
\def\eeqa{\end{eqnarray}}
\newcommand{\no}{\nonumber}
\def\lsim{\mathrel{\rlap{\lower4pt\hbox{\hskip1pt$\sim$}}
    \raise1pt\hbox{$<$}}}         %less than or approx. symbol
\def\gsim{\mathrel{\rlap{\lower4pt\hbox{\hskip1pt$\sim$}}
    \raise1pt\hbox{$>$}}}         %greater than or approx. symbol

\begin{document}

\vspace*{-30mm}

\title{\boldmath Implications of the CDF $t\bar t$ Forward-Backward Asymmetry for Boosted Top Physics}

\author{Kfir Blum}\email{kfir.blum@weizmann.ac.il}
\affiliation{Department of Particle Physics and Astrophysics,
  Weizmann Institute of Science, Rehovot 76100, Israel}

\author{C\'edric Delaunay}\email{cedric.delaunay@weizmann.ac.il}
\affiliation{Department of Particle Physics and Astrophysics,
  Weizmann Institute of Science, Rehovot 76100, Israel}

\author{Oram Gedalia}\email{oram.gedalia@weizmann.ac.il}
\affiliation{Department of Particle Physics and Astrophysics,
  Weizmann Institute of Science, Rehovot 76100, Israel}

\author{Yonit Hochberg}\email{yonit.hochberg@weizmann.ac.il}
\affiliation{Department of Particle Physics and Astrophysics,
  Weizmann Institute of Science, Rehovot 76100, Israel}

\author{Seung J. Lee}\email{seung.lee@weizmann.ac.il}
\affiliation{Department of Particle Physics and Astrophysics,
  Weizmann Institute of Science, Rehovot 76100, Israel}

\author{Yosef Nir}\email{yosef.nir@weizmann.ac.il}
\affiliation{Department of Particle Physics and Astrophysics,
  Weizmann Institute of Science, Rehovot 76100, Israel}

\author{Gilad Perez}\email{gilad.perez@weizmann.ac.il}
\affiliation{Department of Particle Physics and Astrophysics,
  Weizmann Institute of Science, Rehovot 76100, Israel}

\author{Yotam Soreq}\email{yotam.soreq@weizmann.ac.il}
\affiliation{Department of Particle Physics and Astrophysics,
  Weizmann Institute of Science, Rehovot 76100, Israel}

\vspace*{1cm}

\begin{abstract}
New physics at a high scale $\Lambda$ can affect top-related
observables at ${\cal O}(1/\Lambda^2)$ via the interference of
effective four quark operators with the SM amplitude. The
$(\bar u\gamma_\mu\gamma^5 T^a u)(\bar t\gamma^\mu\gamma^5 T^a
t)$ operator modifies the large $M_{t\bar t}$ forward-backward
asymmetry, and can account for the recent CDF measurement. The
$(\bar u\gamma_\mu T^a u)(\bar t\gamma^\mu T^a t)$ operator
modifies the differential cross section, but cannot enhance the
cross section of ultra-massive boosted jets by more than 60\%.
The hint for a larger enhancement from a recent CDF measurement
may not persist future experimental improvements, or may be a
QCD effect that is not accounted for by leading order and
matched Monte Carlo tools or naive factorization.  If it comes
from new physics, it may stem from new light states or an
${\cal
  O}(1/\Lambda^4)$ new physics effect.
\end{abstract}

\maketitle

%%%%%%%%%%%%%%%%%%%%
\mysection{Introduction}
The top quark is unique among the known elementary fermions in
that its coupling to the electroweak symmetry breaking sector
is not small. There is still much to be explored in both the
top quark sector and the electroweak breaking sector. This
situation makes the experimental study of top physics
interesting as a probe of new physics, and promising in its
potential to lead to actual discoveries. The Tevatron
experiments, CDF and D0, are now reaching a stage where the
precision of their top-related measurements might provide first
hints to such new physics.

The CDF Collaboration has recently provided two new intriguing
measurements. First, a large forward-backward $t\bar t$
production asymmetry was observed for large invariant mass of
the $t\bar t$ system~\cite{Aaltonen:2011kc}:
\beq\label{eq:atthexp}
A^{t\bar t}_h\equiv A^{t\bar t}(M_{t\bar t}\geq450\ {\rm
  GeV})=+0.475\pm0.114\,,
\eeq
to be compared with the Standard Model (SM)
prediction~\cite{Almeida:2008ug, Bowen:2005ap,
Antunano:2007da}, $A^{t\bar t}_h=+0.09\pm0.01$.

Second, the CDF Collaboration has recently made progress in
studying the mass distribution of highly boosted jets
($p_T>400$\,GeV for the leading jet)~\cite{cdfboostedqcd}. This
study led to an upper bound of $20$\,fb on the corresponding
boosted top pair cross section, based on naive QCD background
estimation~\cite{cdfboostedtops}. The analysis included two
channels, one involving two massive jets ($130-210$\,GeV) and
another with one massive jet and large missing energy. An
interesting result found was a significant deviation from the
estimated background in the first channel, while no excess was
found in the second channel or in the combined inclusive
search. However, in Ref.~\cite{Eshel:2011vs} it was argued that
the hadronic channel is more sensitive to the presence of
boosted tops, and accounting for the excess in that channel
leads to a tension of less than 1.5 standard deviations in the
missing $E_T$ channel. This observation motivates us to
consider the possibility that the excess is associated with an
enhanced boosted tops cross section, which might also be linked
to Eq.~\eqref{eq:atthexp}.

The estimation of the excess depends on a parameter $R_{\rm
mass}$, described below in Eq.~\eqref{eq:defr}, which
determines the QCD background. Assuming that both the
statistical and systematic uncertainties scale linearly with
$R_{\rm mass}^{-1}$, the cross section for ultra-massive
boosted jets (not coming from QCD events) can be written as
follows
\beq\label{eq:defsigb}
\sigma_b\!\!\equiv \sigma
^{t_h \bar t_h}(p_T\!>\!400\, {\rm GeV})\!\sim \!\left[21\!-
\!\left(8.7\!\pm\!3.1\right) R_{\rm mass}^{-1}\right]
\!{\,\rm fb},
\eeq
where $t_h$ stands for a hadronically decaying top. The SM
prediction is $\sigma^{\rm
SM}_b=2.0\pm0.2$\,fb~\cite{Kidonakis:2003qe}.

It is not unlikely that the differences between either or both of
these measurements and the corresponding SM predictions
will disappear with improved experimental precision, or will be
explained by non-trivial QCD effects. Yet, either or both of
these effects might represent hints for new physics. Our approach in
this work is the following. We interpret the measurement of $A^{t\bar
  t}_h$ in terms of new physics, checking the consistency of such a
scenario with other measurements that do not show any significant
deviation from the SM predictions. Then we extract the predictions of
such new physics explanations for ultra-massive boosted jets at the
Tevatron, and compare to the recent measurement.

Several works have interpreted the recent CDF measurement of
$A^{t\bar t}_h$ within specific models of new
physics~\cite{Cheung:2011qa,Delaunay:2011vv, Cao:2011ew,
Bai:2011ed,Shelton:2011hq,Berger:2011ua,Bhattacherjee:2011nr}.
Similarly, new physics models were
invoked~\cite{Ferrario:2008wm, Ferrario:2009bz,Jung:2009jz,
Cheung:2009ch, Frampton:2009rk, Shu:2009xf,Arhrib:2009hu,
Dorsner:2009mq, Cao:2009uz, Barger:2010mw,Cao:2010zb,
Xiao:2010hm, Chivukula:2010fk, Chen:2010hm,Alvarez:2010js} and
model-independent studies were performed~\cite{Jung:2009pi,
Degrande:2010kt,Jung:2010yn} to explain earlier D0 and CDF
measurements of the inclusive asymmetry~\cite{:2007qb,
Aaltonen:2008hc}.

We do not discuss a specific new physics model, but we focus on
a large class of models with the following two ingredients:
\begin{itemize}
  \item The scale of the new physics is well above the
      scale $M_{t\bar t}$ that is relevant to the CDF
      measurements.
    \item The dominant contribution to $A^{t\bar t}_h$ comes from
      interference between the new physics contribution and the SM
      contribution to top pair production.
    \end{itemize}
These assumptions allow us to follow a low energy model
independent approach, and lead us to particularly clear and
strong conclusions. Ref.~\cite{Degrande:2010kt} has recently
presented a comprehensive analysis of top pair production at
hadron colliders within the same framework. The novelty in our
work is, first, the incorporation of the measurement of the
$M_{t\bar t}$-dependent $A^{t\bar t}$ and, second, the study of
the boosted jets.

%%%%%%%%%%%%%%%%%%%%
\mysection{Boosted jets production}
The CDF study~\cite{cdfboostedtops} focused on events with two
jets, with a lower bound on the transverse momentum ($p_T>400$
GeV) and an upper bound on the pseudorapidity ($\eta<0.7$) of
the leading jet. As concerns the jet masses, CDF has defined
``light'' ($30-50$ GeV) and ``massive'' ($130-210$ GeV) jets.
The search was divided to four regions. Region A corresponds to
events with two light jets, regions B and C to one light and
one massive jet, depending on which is the leading jet in terms
of $p_T$, and region D corresponds to two massive jets. The top
pairs should contribute to region D. To estimate the QCD
contribution to this region, three assumptions were invoked:
\begin{enumerate}
\item Events in regions A,B,C come from only QCD;
\item The actual cross section can be factorized into the partonic
    cross section, which only weakly depends on the masses of the
    final states, and the jet and soft functions;
\item The masses of the leading and sub-leading jets are largely
  uncorrelated.
\end{enumerate}
Under these assumptions,
\beq\label{eq:defr}
R_{\rm mass}\equiv\frac{n_{\rm B} n_{\rm C}}{n_{\rm A} n_{\rm
D}}=1\,,
\eeq
where $n_X$ is the number of QCD events in region $X$.
Assumption 3 above could turn out to be wrong if there is some
mechanism in QCD which leads to bias towards two massive jets.
In~\cite{Eshel:2011vs} it was shown that $R_{\rm mass}$ is
insensitive to the variation of the relative partonic momentum
fraction of the parton distribution function (PDF) value due to
the variation of the jet masses between regions $A$ to $D$.
Furthermore, it is possible to test this assumption by using
various Monte Carlo (MC) tools to extract $R_{\rm mass}$. We
did so with four different tools. The results are summarized in
Table~\ref{Statistic}.

\begin{table}[htdp]
\begin{center}
%\rowcolors{1}{lightgray}{white}
\begin{tabular}{|lcr|}
\hline   {\bf MC tools} & Matching & $R_{\rm mass}$ \\
\hline  Sherpa   & Yes & $0.88\pm0.03$\\
 MadGraph & Yes & $0.86\pm0.04$\\
  MadGraph & No  & $0.76\pm0.04$\\
  Herwig   & No  & $0.86\pm0.02$\\
\hline
\end{tabular}
\end{center}
\caption{The results for $R_{\rm mass}$ [Eq.~(\ref{eq:defr})]
from the different MC tools: Sherpa
(1.2.3)~\cite{Gleisberg:2008ta} with matching (jj,jjj,jjjj),
MadGraph/MadEvent (4.4.56)~\cite{Alwall:2007st} with MLM
matching~\cite{Hoche:2006ph} (jj,jjj,jjjj) to the Pythia
package (2.1.4)~\cite{Sjostrand:2006za}, MadGraph/MadEvent with
no matching, and Herwig (6.520)~\cite{Corcella:2002jc} with no
matching. We use the CTEQ6M PDF set~\cite{Pumplin:2002vw} and
FastJet (2.4.2)~\cite{Cacciari:2005hq} with anti-k$_t$
algorithm~\cite{Cacciari:2008gp} ($\Delta R=1$). Quoted errors
are statistical only.}
\label{Statistic}
\end{table}%

The impressive agreement between Sherpa and MadGraph when
matching is employed leads us to use, instead of
Eq.~(\ref{eq:defr}), the estimate
\beq\label{eq:numr}
R_{\rm mass}=0.87\,.
\eeq
The estimated number of background events within the data
sample of 5.95 fb$^{-1}$ is then
\beqa\label{eq:bkgd}
{\rm QCD}&:& 15\pm5\,,\nonumber\\
t\bar t&:&3\pm1\,.
\eeqa
The number of observed events was 32~\cite{cdfboostedtops},
which constitutes a deviation of $2.7\sigma$ from the above
expectation. Following the exercise performed in
Ref.~\cite{Eshel:2011vs}, the difference between the 32
observed events and the mean value of Eq.~(\ref{eq:bkgd}) is
translated to a cross section of
\beq\label{eq:totexcess}
\sigma_b^{\rm NP}\sim10\pm4\ {\rm fb}\,,
\eeq
or, equivalently,
\beq\label{eq:excess}
N_b\equiv{\sigma^{\rm NP}_b}/ {\sigma^{\rm SM}_b}\sim5\pm2\,.
\eeq
Below we obtain predictions from new physics scenarios for
$N_b$, which we will compare against Eq.~(\ref{eq:excess}).

%%%%%%%%%%%%%%%%%%%%
\mysection{Additional data}
Other top-related CDF and D0 measurements, beyond $A^{t\bar
t}_h$ and $\sigma_b$, do not show significant deviations from
the SM predictions. (Interestingly, a recent D0 measurement of
the differential $p_T$ distribution of $t\bar t$ events hints
towards some increase over the NLO SM prediction for
$p_T\sim300$~GeV~\cite{Abazov:2010js}.) When we invoke new
physics to account for the large value of $A^{t\bar t}_h$, we
will have to make sure that such new physics does not violate
the constraints from other measurements. Specifically, we
consider the following measurements:

(i) The forward-backward $t\bar t$ production asymmetry for
small invariant mass of the $t\bar t$
system~\cite{Aaltonen:2011kc}:
\beq\label{eq:attlexp}
A^{t\bar t}_l\equiv A^{t\bar t}(M_{t\bar t}\leq450\ {\rm
  GeV})=-0.116\pm0.153\,,
\eeq
to be compared with the SM
prediction~\cite{Almeida:2008ug}, $A^{t\bar t}_l=+0.040
\pm0.006$.

(ii) The inclusive $t\bar t$ production cross section reported
by the CDF Collaboration~\cite{cdfcrosssection,
Schwanenberger:2010un}:
\beq\label{eq:totsig}
\sigma_i\equiv\sigma^{t\bar t}_{\rm inclusive}=7.50\pm0.48\
{\rm pb}\,,
\eeq
which is consistent with the D0 result~\cite{d0crosssection}.
This is to be compared with the SM
prediction~\cite{Kidonakis:2010dk}, $\sigma_i=7.2\pm0.4$ pb. We
note that the results of~\cite{Kidonakis:2010dk} agree with
other recent evaluations~\cite{Cacciari:2008zb,
Langenfeld:2009wd}, but are in some tension
with~\cite{Ahrens:2010zv}.  We conservatively use this result,
as that of~\cite{Ahrens:2010zv} would be less constraining
given our framework.

(iii) The $t\bar t$ differential cross section, which for
simplicity we choose to represent by the following large
$M_{t\bar t}$ bin~\cite{Aaltonen:2009iz}:
\beq\label{eq:higsig}
\sigma_h\equiv \sigma^{t\bar t}(700\ {\rm GeV}<M_{t\bar
  t}<800\ {\rm GeV})=80\pm37\ {\rm fb}\,,
\eeq
to be compared with the SM
prediction~\cite{Almeida:2008ug,Ahrens:2010zv},
$\sigma_h=80\pm8$ fb. The choice of this specific bin requires
some explanation.
\begin{itemize}
\item Since we focus on new physics which contributes to
    the $t\bar t$ cross section $\propto(M_{t\bar
    t}/\Lambda)^2$ relative to the SM, the corrections to
    lower $M_{t\bar t}$ bins are less significant.
\item In the more recent study of~\cite{Aaltonen:2011kc},
    which was based on a larger sample, there is some
    discrepancy above 800~GeV (note however that the data
    in~\cite{Aaltonen:2011kc} is not unfolded to the
    partonic level and so cannot be directly used). Hence
    we choose to use the next-to-last bin given
    in~\cite{Aaltonen:2009iz}.
\end{itemize}

In order to minimize the impact of NLO corrections to the new physics
(NP) contributions, we normalize the new physics contribution to the
SM one. We assume that the $K$-factors are universal, so that the
NP/SM ratios at LO and NLO are the same.  Since the highly virtual
intermediate gluon in the SM process can be integrated out to give
${\cal O}_V^8\,$, NP NLO contributions should be similar, up to small
corrections of ${\cal O}(\alpha_s)$.  Moreover, the parity invariance
of QCD suggests that the same argument applies to ${\cal O}_A^8\,$ as
well.

Combining in quadrature the experimental and theoretical
uncertainties, we represent Eqs.~(\ref{eq:totsig})
and~(\ref{eq:higsig}) as follows:
\beqa\label{eq:ninh}
N_i\equiv \left| \sigma_i^{\rm NP}/\sigma_i^{\rm SM} \right| &\lesssim&0.1\,,\nonumber\\
N_h\equiv \left| \sigma^{\rm NP}_h/\sigma^{\rm SM}_h \right|
&\lesssim&0.5\,.
\eeqa
%

%%%%%%%%%%%%%%%%%%%%
\mysection{${\cal L}_{\rm eff}$ for $t\bar t$ production}
The basic assumption that we aim to test is that the source of the
large value of $A^{t\bar t}_h$ is new physics that is characterized by
a mass scale $\Lambda$ that is larger than $M_{t\bar t}$ in all the
measurements that we consider. (In particular, our Tevatron-related
calculations are safe for $\Lambda\gg1$ TeV.) In such a case, the new
physics can be represented as a set of effective operators. These
operators must lead from an initial $u\bar u$ state to a final $t\bar
t$ state. (The contribution of $d\bar d\to t\bar t$ at the Tevatron is
at most $15\%$ that of $u\bar u\to t\bar t$ for $M_{t\bar t}$ above
450~GeV, as relevant for the observables that we consider.) When
expanding in inverse powers of the scale $\Lambda$, the leading NP
contributions to top pair production appear at ${\cal
  O}(1/\Lambda^2)$:
\beq
|M|^2=|M_{\rm SM}|^2+2{\cal R}e(M_{\rm SM}M^*_{\rm NP})+{\cal
  O}(1/\Lambda^4)\,.
\eeq
Therefore, we should consider dimension-six operators that
interfere with the SM amplitude. There are two such four-quark
operators:
\beqa\label{eq:basis}
{\cal L}_{\rm eff}^{4q}&=&\frac{1}{\Lambda^2}\left(c_A^8 {\cal
O}_A^8+c_V^8
  {\cal O}_V^8\right)\,,\nonumber\\
{\cal O}_A^{8}&=&(\bar u\gamma_\mu\gamma^5 T^a u)
(\bar t\gamma^\mu\gamma^5 T^a t)\,,\nonumber\\
{\cal O}_V^{8}&=&(\bar u\gamma_\mu T^a u) (\bar t\gamma^\mu T^a
t)\,.
\eeqa
Below, we consider the effects of these two operators on the
forward-backward asymmetry and on the differential cross
section in top pair production. We work only at leading order,
using the MSTW PDF set~\cite{Martin:2009iq} and running of the
strong coupling at leading order. We use factorization and
renormalization scales given by the partonic center of mass
energy. Note that all other possible Lorentz structures
(scalar, pseudoscalar, tensor and pseudotensor) and the other
possible color contractions do not interfere with the SM
amplitude.

In addition to the four-quark operators, there is a
chromomagnetic dipole operator,
\beq\label{eq:chma}
{\cal L}_{\rm eff}^{tg}=\frac{c_{tg} v}{\Lambda^2}(\bar
t\sigma_{\mu\nu} T^a t)G^{a\mu\nu}\,.
\eeq
Here $v$ is the vacuum expectation value of the Higgs field,
reflecting the fact that the operator breaks $SU(2)$. The
corresponding chromoelectric dipole operator violates CP and,
therefore, does not interfere with the SM amplitude. The
interference of the chromomagnetic operator requires a
chirality flip. Consequently, the corresponding operator
involving the up quark is suppressed by $m_u$ and therefore
negligible. Thus, among the dipole operators, Eq.~(\ref{eq:chma})
is the only one that we need to consider.

The interference of the $c_{tg}$ term with the SM amplitude
does not contribute to the forward-backward asymmetry. As
concerns the contribution to the cross section, it falls like
$1/M_{t\bar t}^2$~\cite{Degrande:2010kt}. We learn that while
the $c_{tg}$ term can affect the inclusive cross section, it
does not affect $A^{t\bar t}_h$, and its effect on $N_h$ and
$N_b$ is negligible. We therefore focus mainly on the effects of
${\cal O}_A^8$ and ${\cal O}_V^8$. See, however, additional
discussion above Eq. (\ref{eq:nbsmall}).

%%%%%%%%%%%%%%%%%%%%
\mysection{The forward-backward asymmetry}
It is convenient to represent the new physics effects on
$A^{t\bar t}$ as follows:
\beq
(A^{t\bar t})^{\rm NP}=\frac{\sigma_-^{\rm NP}}{\sigma_+^{\rm
SM}+\sigma_+^{\rm NP}}\,,
\eeq
where $\sigma_\pm\equiv\sigma(\Delta y>0)\pm\sigma(\Delta y<0)$
and $\Delta y$ is the rapidity difference, $\Delta
y=y_t-y_{\bar t}$. Among the two operators of
Eq.~(\ref{eq:basis}), only ${\cal O}_A^{8}$ contributes to
$\sigma_-\,$. If this is the only NP operator, the NP
contribution to $A^{t\bar t}_h$ is
\beq
(A^{t\bar t}_h)^{\rm NP}\simeq0.17\frac{c_A^8}{\Lambda^2_{\rm
TeV}}\,,
\eeq
where $\Lambda_{\rm TeV}=\Lambda/{\rm TeV}$. Requiring that
$(A^{t\bar t}_h)^{\rm NP}\sim+0.4\pm0.1$, we obtain
\beq\label{eq:aaafb}
{c_A^8}/{\Lambda^2_{\rm TeV}}\sim2.4\pm0.7\,.
\eeq
Eq.~(\ref{eq:aaafb}) implies, in turn,
\beq
(A^{t\bar t}_l)^{\rm NP}\sim+0.10\pm0.03\ \Longrightarrow\
A^{t\bar t}_l=+0.14\pm0.04\,,
\eeq
about $1.7\sigma$ higher than the experimental result in
Eq.~(\ref{eq:attlexp}). In addition, Eq.~(\ref{eq:aaafb})
predicts $(A^{t\bar t})^{\rm NP}\sim+0.21\pm0.06$, $1.5\sigma$
too large~\cite{Aaltonen:2011kc}, and $(A^{t\bar t}(\Delta
y>1))^{\rm NP}\sim+0.55\pm0.15$, within one standard deviation
from the measurement~\cite{Aaltonen:2011kc}. On the other hand,
the ${\cal O}_A^8$ operator does not affect the cross section
at ${\cal O}(1/\Lambda^2)$ and, in particular, cannot enhance
the boosted jets cross section. The contribution of the next
order in $1/\Lambda^2$ to the forward-backward asymmetry is
subdominant, and, using the one sigma lower bound of
Eq.~\eqref{eq:aaafb}, saturates the constraint from $N_h$ in
Eq.~\eqref{eq:ninh}.

Eq.~(\ref{eq:aaafb}) provides an upper bound on the scale of
new physics. We use naive dimensional analysis (NDA) to derive
an upper bound on $c_A^8$,
\beq
c_A^8\lesssim 16\pi^2\,.
\eeq
Combining this upper bound with the one sigma lower bound in
Eq.~(\ref{eq:aaafb}), we obtain
\beq\label{eq:afbscale}
\Lambda\lesssim 10\ {\rm TeV}\,.
\eeq

The upper bound on $\Lambda$ in Eq.~(\ref{eq:afbscale}) implies
that, if a heavy axigluon is to provide a perturbative
explanation to the large asymmetry in Eq.~(\ref{eq:atthexp}),
then new physics effects should be observed early on at the
LHC. In particular, given that the LHC will directly explore
energy scales close to $\Lambda\,$, then the $t\bar t$
production cross section should be significantly enhanced at
high $M_{t\bar t}$ (see~\cite{future} for more details).

\begin{figure}[hbp]
\includegraphics[width=0.45\textwidth]{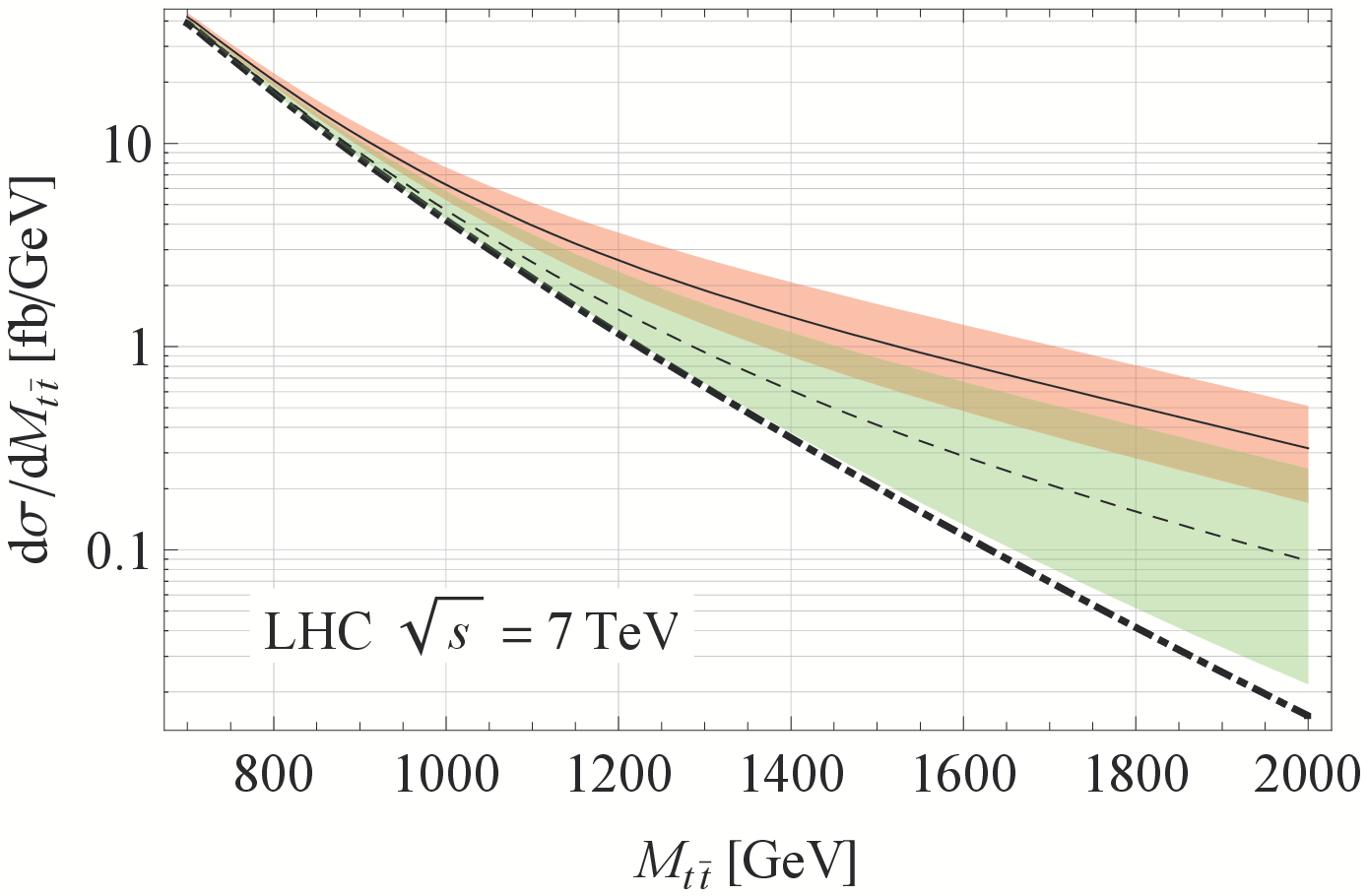}\\
\includegraphics[width=0.45\textwidth]{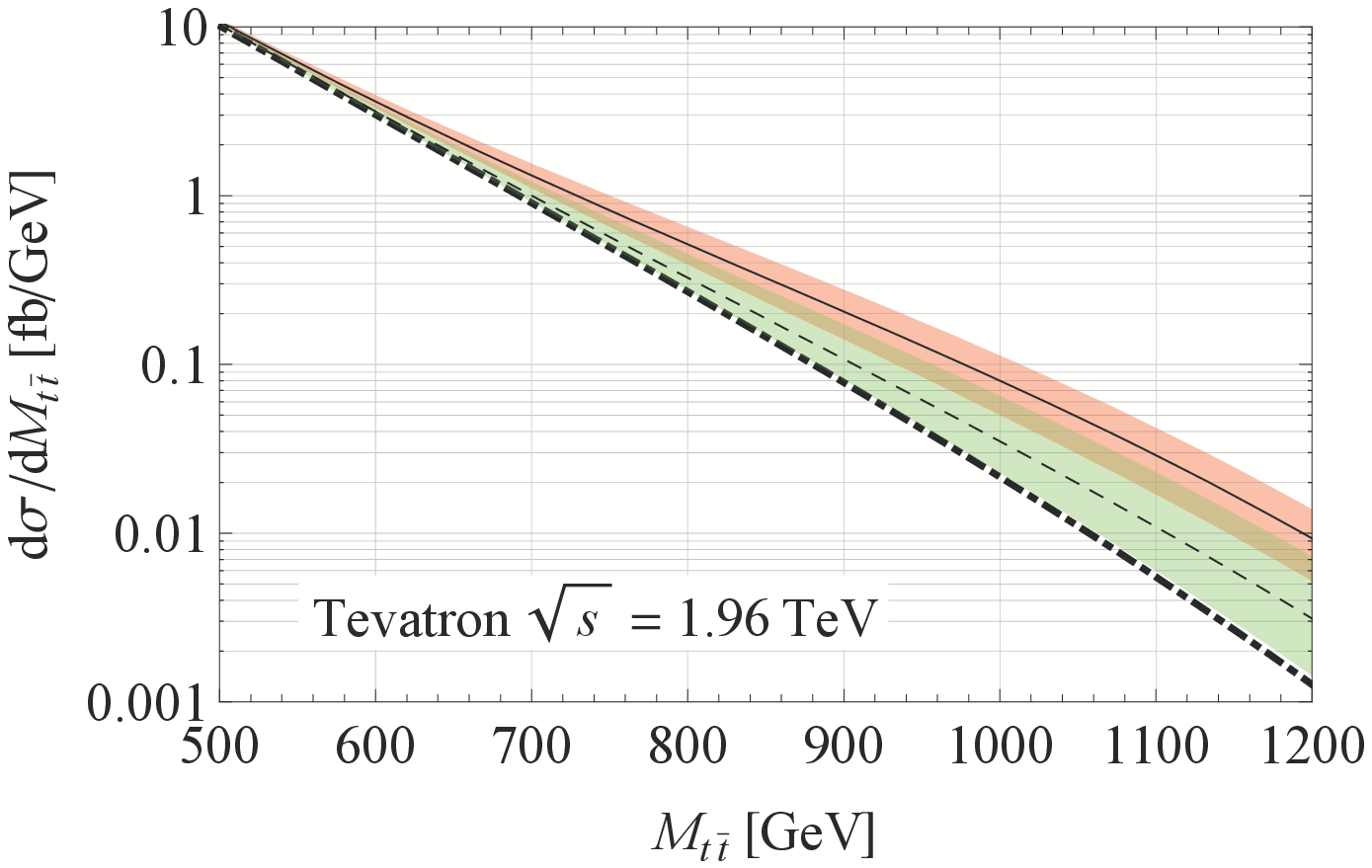}
\caption{The differential cross section of top pair production as a
  function of $M_{t\bar t}$ at the LHC at 7 TeV (top) and the Tevatron (bottom), calculated at leading
  order. The dashed-dotted curve corresponds to $\sigma^{\rm SM}_{\rm LO}$. The
  solid curve corresponds to $\sigma^{{\rm SM}+{\cal O}_A^{8}}$, where the
  new physics coupling is set by the central value of $A^{t\bar t}_h$. The dashed
  curve corresponds to $\sigma^{{\rm SM}+{\cal O}_A^{8}}$, where the new physics
  coupling is set by the central value of the inclusive asymmetry $A^{t\bar t}$.
  The shaded regions around the two upper curves depict the one sigma ranges of the
  corresponding measurements.}
\label{fig:lhccs}
\end{figure}

\begin{figure}[hbp]
\includegraphics[width=0.45\textwidth]{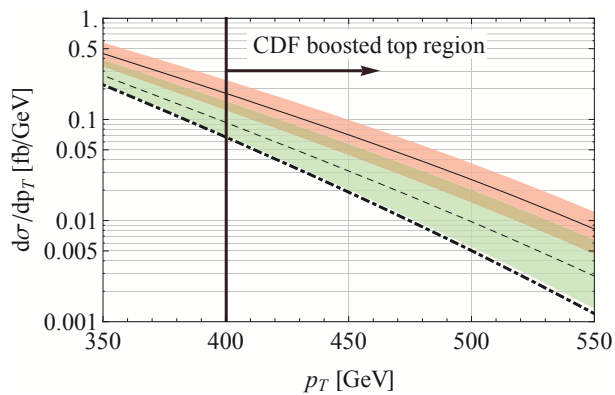}
\caption{The differential cross section of top pair production as a
  function of $p_T$ at the Tevatron, calculated at leading
  order. The color and curve conventions are the same as in Fig.~\ref{fig:lhccs}.
  The vertical line corresponds to the lower $p_T$ cut used in the analysis of~\cite{cdfboostedtops}.}
\label{fig:tevpt}
\end{figure}

To substantiate this statement, we perform the following
exercise. We note that the ${\cal O}_A^8$ operator does modify
the cross section at ${\cal O}(1/\Lambda^4)$ via the $|M_{\rm
NP}|^2$ term. We plot in Fig.~\ref{fig:lhccs} the differential
$t\bar t$ cross section as a function of $M_{t\bar t}$ at the
LHC for the case where the SM is augmented by only the ${\cal
O}_A^{8}$ operator, with the coupling of Eq.~(\ref{eq:aaafb})
(the distribution at the Tevatron is also depicted for
comparison). Of course, at this order there are many more
operators that affect the cross section, either via $|M_{\rm
NP}|^2$ for ${\cal O}(1/\Lambda^2)$ operators, or via ${\cal
R}e(M_{\rm SM}M_{\rm NP}^*)$ for ${\cal O}(1/\Lambda^4)$
operators. In Ref.~\cite{future} it is shown, however, that
there can be no fine-tuned cancellations between these
other contributions and the one that we consider. Thus our
calculation illustrates the size of the effects that should be
expected at the LHC. We learn that at $M_{t\bar t}\sim1.5$ TeV,
we should expect an enhancement by a factor $\sim5$ compared to
the SM. When applied to the Tevatron, the same exercise gives
an enhanced boosted jets production cross section of $N_b\sim
2$, which is $1.5\sigma$ from the mean value in
Eq.~\eqref{eq:excess}. Fig.~\ref{fig:tevpt} depicts the
resulting $p_T$ distribution at the Tevatron.

We stress that the recent measurement of the differential
$t\bar t$ forward-backward asymmetry predicts a more pronounced
deviation from the SM at the LHC than the previous inclusive
asymmetry measurement. This is illustrated in
Fig.~\ref{fig:lhccs} by the difference between the solid and
dashed curves (and the respective shaded regions) and the
dashed-dotted curve representing the SM prediction.

%%%%%%%%%%%%%%%%%%%%
\mysection{The $t\bar t$ production cross section}
Among the two operators of Eq.~(\ref{eq:basis}), only ${\cal
O}_V^{8}$ contributes to the inclusive cross section
($\sigma_i$), to the cross section at large $M_{t\bar t}$
($\sigma_h)$ and to the production cross section of boosted
jets ($\sigma_b)$ at ${\cal O}(1/\Lambda^2)$:
\beqa
N_i&\simeq&0.24\ {c_V^8}/{\Lambda^2_{\rm TeV}}\,,\no\\
N_h&\simeq&0.76\ {c_V^8}/{\Lambda^2_{\rm TeV}}\,,\no\\
N_b&\simeq&1.5\ {c_V^8}/{\Lambda^2_{\rm TeV}}\,.
\eeqa
This equation, where relevant, is consistent with previous results in
the literature (\cite{Degrande:2010kt} and references therein).
Independently of the value of the coupling, our framework predicts
\beq
N_b\sim 2 N_h \sim 6N_i\,.
\eeq
This ordering of the size of the effects reflects the fact that
each of these three measurements samples a different $M_{t\bar
t}$ region; the closer this region is to $\Lambda$, the larger
the effect.

The relation between $N_b$ and $N_i$ and between $N_h$ and
$N_i$ can be modified by the presence of the chromomagnetic
dipole operator in Eq.~\eqref{eq:chma}, if $\left| c_{tg}
\right|$ is not much smaller than $\left|c_V^8 \right|$.
However, since the $c_{tg}$ term does not affect the cross
section at high invariant mass $M_{t \bar t}^2\gg m_t^2$, the
relation between $N_b$ and $N_h$ is insensitive to it. The
bound on $N_h$ in Eq.~\eqref{eq:ninh} then leads to an upper
bound on the enhancement of boosted jets production:
\beq\label{eq:nbsmall}
N_b\lesssim0.8\,,
\eeq
well below our estimate of Eq.~(\ref{eq:excess}). We conclude
that one of the following must hold:
\begin{itemize}
  \item The estimate of Eq.~(\ref{eq:excess}) is wrong
      because of either experimental or QCD effects.
  \item New physics explains Eq.~(\ref{eq:excess}), but it is
      characterized by a scale that is $\lesssim1$ TeV.
  \item Heavy new physics explains Eq.~(\ref{eq:excess}), but
      ${\cal O}(1/\Lambda^4)$ terms play an important
      role~\cite{future}.
  \item The reported excess in events with two boosted
      massive jets does not originate from top quarks.
\end{itemize}

%%%%%%%%%%%%%%%%%%%%
\mysection{Conclusions}
The recent CDF measurement of the $t\bar t$ forward-backward
asymmetry at large $M_{t\bar t}$, $A^{t\bar t}_h$, shows a
deviation higher than $3\sigma$ from the SM prediction. The
recent CDF measurement of ultra-massive boosted jets,
$\sigma_b$, shows a deviation of order $2.7\sigma$ from a SM
calculation augmented with an estimate of QCD background based
on data and on several simplifying assumptions that we test
with various MC tools.

We investigated whether these effects can be accounted for
within a large class of new physics models. This class of
models is defined by a mass scale above the scales directly
explored by these CDF measurements, and a dominant effect
coming from interference between the Standard Model and new
physics amplitudes.

Within this framework, we find that there is a single four
quark operator that contributes to the asymmetry, which is the
axial vector, color octet, operator ${\cal O}_A^8=(\bar
u\gamma_\mu\gamma^5 T^a u) (\bar t\gamma^\mu\gamma^5 T^a t)$.
There is a different single four quark operator that modifies
the differential cross section at high $t\bar t$ invariant
mass, which is the vector, color octet, operator ${\cal
O}_V^8=(\bar u\gamma_\mu T^a u) (\bar t\gamma^\mu T^a t)$. This
means in particular that there is no model independent relation
between the forward-backward asymmetry and the boosted jets
cross section. Note that we focus on these operators at tree
level, and so do not discuss their contribution to dijet
production at the LHC~\cite{Bai:2011ed}.

Our numerical results are summarized in
Table~\ref{Observables}. If ${\cal O}_A^8$ accounts for the
high value of $A^{t\bar t}_h$, then the asymmetry at low
invariant mass is about $1.7\sigma$ high compared to the CDF
measurement. One should expect a striking enhancement of $t\bar
t$ production at high $M_{t\bar t}$ at the LHC.

\begin{table}[tbdp]
\begin{center}
%\rowcolors{1}{lightgray}{white}
\begin{tabular}{|lcccr|}
\hline   Obs. & Def. & Experiment & Standard Model & New physics \\
\hline  $A^{t\bar t}_h$  & Eq.~(\ref{eq:atthexp}) & $+0.475\pm0.114$ &
$+0.09\pm0.01$ & Input \\
  $A^{t\bar t}_l$  & Eq.~(\ref{eq:attlexp}) & $-0.116\pm0.153$ &
$+0.040\pm0.006$ & $+0.16\pm0.04$ \\
  $\sigma_i$ & Eq.~(\ref{eq:totsig}) &  $7.50\pm0.48$ pb &
$7.2\pm0.4$ pb & Input \\
  $\sigma_h$ & Eq.~(\ref{eq:higsig}) &  $80\pm37$ fb &
$80\pm8$ fb & Input  \\
  $\sigma_b$ & Eqs.~(\ref{eq:defsigb},\ref{eq:numr}) & $12\pm4$ fb &
$2.0\pm0.2$ fb & $<3.2$ fb  \\
\hline
\end{tabular}
\end{center}
\caption{Effects from new physics of ${\cal O}(1/\Lambda^2)$ on
  top-related observables. The first column gives the list of
  observables, and the second the equation where they are defined. We
  use $A^{t\bar t}_h$, $\sigma_i$ and $\sigma_h$ to fix, or constrain,
  the new physics parameters. The experimental value quoted for
  $\sigma_b$ is based on our theoretical interpretation of the data.}
\label{Observables}
\end{table}%

If ${\cal O}_V^8$ is to be consistent with constraints from the
inclusive and differential cross sections, then it cannot
enhance the boosted tops cross section by more than $60$\%.
Furthermore, ${\cal O}_V^8$ is restricted to be significantly
smaller than the contribution of ${\cal O}_A^8$ implied by the
$t\bar t$ asymmetry. This means that a chiral model cannot
consistently reproduce the asymmetry.

The above conclusions are related to the fact that the
interference effects of heavy new physics with the SM scale
roughly as $(M_{t\bar t}/\Lambda)^2$ relative to the SM.
Consequently, they do not differentiate between the low and
high $M_{t\bar t}$ regions enough to avoid tension with the
data.

The conclusion concerning the ultra-massive boosted tops is
that ${\cal O}(1/\Lambda^2)$ effects do not explain the
discrepancy of the data with our theoretical estimate of the SM
contribution. Perhaps the explanation does not involve new
physics: The deviation is below $3\sigma$ and might disappear
with better experimental accuracy, or it could be that QCD
effects that are unaccounted for in the various MC tools play a
role. If the deviation is related to new physics, then either
the new physics is below the TeV scale and cannot be
represented by effective higher-dimension operators, or the
contribution of $|M_{\rm NP}|^2\propto1/\Lambda^4$ is
significant, bringing into the analysis a richer set of
operators and a sharper distinction between the low and high
$M_{t\bar t}$ regions.

The LHC will explore $t\bar t$ production at higher energy
scales. Whether the scale $\Lambda$ is within its direct reach
or just beyond it, new physics effects are expected to be
large. The Tevatron, on the other hand, has better access to
the $q\bar q\to t\bar t$ process which, via observables such as
the forward-backward asymmetry, can close in on the detailed
structure of new physics. The combination of Tevatron and LHC
measurements is likely to shed light on the top-related puzzles
very soon.

%%%%%%%%%%%
\mysections{Acknowledgments} We thank Raz Alon, Yochay Eshel,
Alex Kagan, Fabrizio Margaroli, Pekka Sinervo and George
Sterman for useful discussions. YN is the Amos de-Shalit chair
of theoretical physics and supported by the Israel Science
Foundation (grant \#377/07), the German-Israeli foundation for
scientific research and development (GIF), and the United
States-Israel Binational Science Foundation (BSF). GP is the
Shlomo and Michla Tomarin career development chair and is
supported by the Israel Science Foundation (grant \#1087/09),
EU-FP7 Marie Curie, IRG fellowship, Minerva and G.I.F., the
German-Israeli Foundations, and the Peter \& Patricia Gruber
Award.

%%%%%%%%%%%%%%%%%%%%%

\end{document}